# SWCNT@BNNT with 1D van der Waals Heterostructure with a High Optical Damage Threshold for Laser Mode-locking

Zheyuan Zhang,[1] Pengtao Yuan,[1] Shoko Yokokawa,[1] Yongjia Zheng,[2] Anton S. Anisimov,[3] Esko I. Kauppinen,[4] Rong Xiang,[2] Lei Jin,[1] Shigeo Maruyama,[2,5] Shinji Yamashita,[1] Sze Yun Set[1, *]

[1] Research Center for Advanced Science and Technology, The University of Tokyo, 4-6-1 Komaba, Meguro-ku, Tokyo 153-8904, Japan
[2] Departure of Mechanical Engineering, The University of Tokyo, 7-3 Hongo, Bunkyo-ku, Tokyo 113-8656, Japan
[3] Canatu, Ltd., Konalankuja 5, FI-00390 Helsinki, Finland
[4] Department of Applied Physics, Aalto University School of Science, 15100, FI-00076 Aalto, Finland
[5] Energy NanoEngineering Lab., National Institute of Advanced Industrial Science and Technology (AIST), 1-2-1 Namiki, Tsukuba, 305-8564, Japan
*Corresponding author: set@cntp.t.u-tokyo.ac.jp

**Abstract:** Single-walled carbon nanotube encapsulated in boron nitrite nanotube (SWCNT@BNNT) is a novel nanomaterial with a one-dimensional van der Waals (1D-vdW) heterostructure. In this paper, we demonstrated that the SWCNT@BNNT has an enhanced optical power tolerance compared to that of the pristine SWCNT while exhibiting an optical saturable absorption properties. Under optical power intensity of 13kW/cm$^2$, the lifetime of the SWCNT@BNNT is found to be 2,270 times longer than SWCNT for 1% degradation in absorbance. A short-cavity mode-locked laser with a high repetition rate of 1 GHz has been realized using the SWCNT@BNNT as the saturable absorber. We have shown that the technique of fabricating nanomaterial with a 1D-vdW heterostructure can modify and enhance the optical properties of the encapsulated nanomaterials.

## 1. Introduction

Carbon nanotube (CNT) is an allotrope of carbon that is made of carbon atoms arranged in a form of cylindrical hexagonal lattice. Since its discovery in 1991 [1], it has been found useful in a wide range of applications because of its unique mechanical [2], electrical and [3, 4] thermal properties [5]. CNT is also a promising material for various optical applications such as fast photovoltaic devices [6], nano-scale light emitters [7] and photo-detectors [8]. Single-walled carbon nanotube (SWCNT) is one of the most reliable and flexible materials as an ultrafast optical saturable absorber for the generation of ultrafast pulses in the picosecond and femtosecond regime [9-17]. Optical saturable absorption is an optical nonlinear phenomenon where the high intensity light "saturates" the optical absorption of the material, resulting in a reduced absorption for optical pulses with a high peak power. It is the key component for passive mode-locking of lasers, to suppress the continuous-wave light and to favor the generation ultrashort pulses. However, in term of heat resistance, the SWCNT has a moderate thermal threshold for oxidization in air ~400 °C [25], which poses a limit to the maximum output optical power. The peak optical power damage threshold is around 200 MW/cm$^2$ (with an average optical power damage threshold of 4.4 kW/cm$^2$) and when it is coated on the facet of the core section of a standard single-mode optical fiber, the resulting maximum optical power damage threshold is merely 4 mW [18]. Although the optical damage of SWCNT due to thermal oxidation can be mitigated by hermetic sealing in a nitrogen atmosphere, resulting in a much higher optical damage threshold to withstand an intracavity power >300 mW [20], it is costly and inconvenient. In order to realize mode-locked lasers with high power levels, there is a strong need for new materials with a higher optical damage threshold and an ultrafast saturable properties similar to that offered by SWCNT.

Van der Waals (vdW) heterostructure is a concept in nanomaterial synthesis that combine atomic layers of different elements to achieve "property by design" [22, 23]. In such a structure, atomic layers are stacked on each other, and different ingredients can be combined beyond symmetry and lattice matching, enabling fabrication of sophisticated structures layer by layer. This concept had been limited in two dimensional (2D) materials like graphene, and were not available for one-dimensional (1D) materials like SWCNT, until recently the technique of a 1D vdW heterostructure was proposed [24]. This technique marks a starting point for a series of function-designable semiconducting nanotube materials with 1D heterostructures that may combine 1D nanomaterials such as SWCNTs, boron nitride nanotubes (BNNTs) and other semiconducting nanotubes.

## 2. SWCNT@BNNT

Single-walled carbon nanotube encapsulated in boron nitride nanotube (SWCNT@BNNT) is a typical 1D vdW nanomaterial, in which the SWCNT is serving as a baseplate and BNNT is grown on it coaxially like a cladding (Fig. 1). The fabrication of the SWCNT@BNNT starts from a thin layer of pristine SWCNT film synthesized by aerosol chemical vapor deposition (CVD) [26], followed by a second CVD growth of BN layer on the SWCNT in an electric furnace. In the furnace, ammonia borane ($H_3NBH_3$) is heated to 70-90 °C at the upstream as BN precursor, and carried by argon gas to a low-pressure thermal CVD chamber at a rate of 300 sccm under a temperature of 1000-1100 °C and a pressure of 300 Pa. The BNNT will start to grow from the open-end of based nanotube and may take several hours to cover the outer surfaces evenly on the entire SWCNT [27]. The number of the external BNNT walls synthesized on the SWCNT can be controlled by an appropriate adjustment of the CVD growth time.

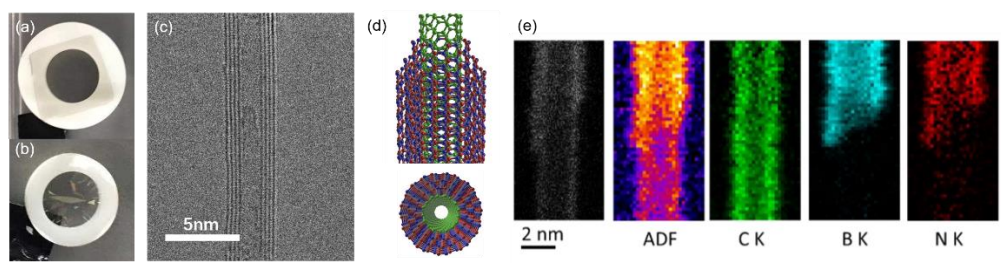

**Fig. 1.** Picture of (a) pristine SWCNT membrane and (b) SWCNT@BNNT membrane. (c) TEM image and (d) atomic models of a SWCNT wrapped with two layers of BNNT (green nodes: carbon atoms; red nodes: nitrogen atoms; blue nodes: boron atoms). (e) Electron energy loss spectroscopic (EELS) mapping of a SWCNT partially wrapped with BNNT, showing that the inner part is carbon and outer layer is BN [24].

Several tests have been performed on the fabricated SWCNT@BNNT to investigate its properties and structure. The X-ray photoelectron spectroscopy (XPS) result reveals only B-N bond and C-C bond but no apparent sign of C-N and B-C, suggesting that there is no noticeable substitution of atoms. The SWCNT structure is well preserved in the BN coating process without interacting with the BN layer. It is expected that the optical properties of SWCNT such as saturable absorption will also be preserved after the BN coating process. This is because BNNT has a wide band gap of ~5.5 eV, which is much larger than that of the SWCNT (~1 eV) where saturable absorption occurs [28, 29]. Extra conduction-band/valence-band will not be introduced to the band structure of the SWCNT. As shown in Fig. 2, the optical absorption spectra of pristine SWCNT and SWCNT@BNNT are almost the same in the infrared region, except for an emergence of an absorption peak near 205 nm which increases with the number of outer BNNT layers. There is also a slight shift of the 3 absorption peaks ($M_{11}$, $S_{22}$, $S_{11}$) towards a longer wavelength. The BN coating does not seem to have altered the chirality distribution and the low-energy-level band distribution of the inner SWCNT by much.

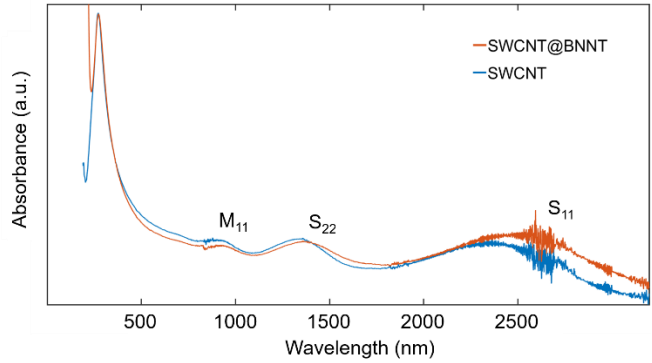

**Fig. 2.** Optical absorption spectra of pristine SWCNT (blue) and SWCNT@BNNT (orange), measured using a spectral photometer.

## 3. Experimental Setup

Fig. 3 shows the experimental setup used for the evaluation of the optical properties of the pristine SWCNT and the SWCNT@BNNT. A mode-locked pulsed fiber laser operating at 1560 nm with an average output power of 4 mW, a pulse-width of 800 fs and a repetition rate of 50 MHz is used as the light source. 25% of the laser output power is tapped and measured with a power meter (PM2) as the reference power. The remaining 75% of the laser output power is amplified by an erbium-doped fiber amplifier (EDFA) to a fixed average power of 45 mW with an estimated peak power at 1 kW, a power level limited by self-phase-modulation induced spectral broadening in single-mode fiber. After amplification, the laser optical power is controlled using an electrically-programmable-variable optical attenuator (VOA), instead of. pump power control of the EDFA. This will ensure a constant optical spectral profile of the test laser source to minimize the influence of wavelength dependence. The attenuated light from the EDFA is launched to free-space via a fiber collimator with a beam diameter of 3.6 mm, and then focused with an aspheric lens (focal length = 18.4 mm) onto the test sample. The transmitted light through the sample is collected by another power meter (PM1) on the axis of the collimated light. The transmittance of the sample can then be measured and calculated. An open-aperture Z-scan measurement technique [30] is employed here, where the optical intensity on the sample can be changed with the sample position, for better understanding of the SWCNT and SWCNT@BNNT.

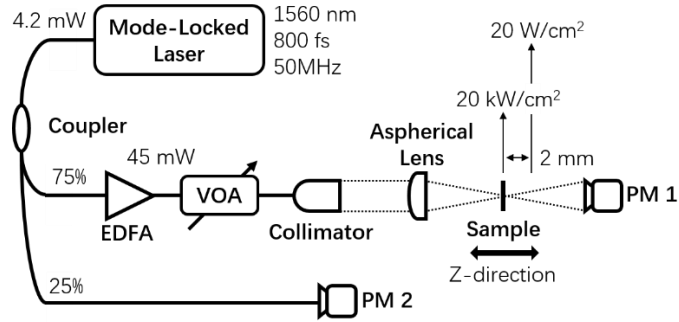

**Fig. 3.** Experimental setup of the saturable absorption measurement (EDFA: Erbium-doped fiber amplifier, VOA: variable optical attenuator, PM: power meter, SUT: sample under test).

## 4. Results and Analysis

The measured total insertion loss of the pristine SWCNT and SWCNT@BNNT are 10% and 19%, respectively. In the Z-scan measurement, the average optical power from the VOA is set to 5 mW, corresponding power intensity of 2 kW/cm² at the focal point. As shown in Fig. 4(a), when the sample is moved in the z-axis across the focal point with the minimum beam diameter, the transmittance maximizes at the focal point (z = 0 mm), a typical signature of a saturable absorption response. It is worth noting that although the transmittance peak of the SWCNT@BNNT (orange traces) is as twice the height as that of the pristine SWCNT (blue traces), it is the absorption that define the modulation depth of saturable absorption. The saturable absorption modulation depth, defined as the intensity-dependent nonlinear absorption divided by the non-saturable absorption, of the SWCNT@BNNT and the pristine SWCNT are measured and estimated to be 2.72% and 2.935%, as shown in Fig. 4(b). The modulation depths of the two samples are almost identical, indicating that the saturable absorption property of the SWCNT has been preserved after the BN coating process.

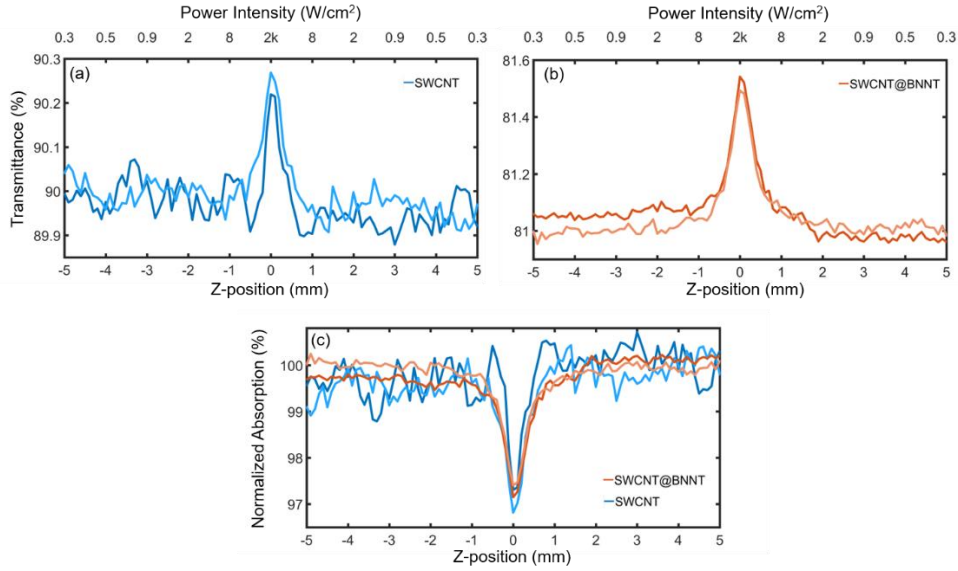

**Fig. 4.** Transmittance of (a) pristine SWCNT and (b) SWCNT@BNNT and (c) corresponding normalized absorption of pristine SWCNT (blue lines) and SWCNT@BNNT (orange lines) in Z-scan experiments.

Since the BNNT layer has a much higher resistance to thermal oxidation up to ~ 900 °C, which is much higher than that of the SWCNT of ~ 400 °C [25], it is expected that the SWCNT@BNNT would possess better thermal properties compared to that of the pristine SWCNT. It is estimated that the growth of BN coating on SWCNT raises the thermal durability of the pristine SWCNT film from ~ 400 °C to an elevated temperature as high as 700 °C [24]. Accordingly, a higher optical damage threshold can also be expected from the SWCNT@BNNT due to the enhancement of its thermal durability.

Using the setup in Fig. 3, the thermal resilience (or heat tolerance) properties of the pristine SWCNT and that of the SWCNT@BNNT are also investigated. The sample is placed at the focal point of the aspherical lens where the optical power intensity is maximized for 10 minutes in order to induce permanent optical damage due to the heat generated by the intense light. No obvious change in transmittance of SWCNT@BNNT has been observed under optical intensity of 10 kW/cm², so the laser average power is set to 30 mW (13 kW/cm²). The changes in the absorbance of SWCNT and SWCNT@BNNT are shown in Fig. 5 (a) and (b), respectively. It is obvious that the absorbance of the pristine SWCNT decreases far more quickly than that of the SWCNT@BNNT, because of graphitization and oxidation caused by the intense accumulated heat. The absorbance of the pristine SWCNT is decreased by about 0.8% within 10 minutes of illumination, while the change in the absorbance of the SWCNT@BNNT is only 0.2%. A logarithmic fitting has been made with the measured traces, resulting in:

$$\Delta A_{SWCNT} \propto 0.0044 \cdot \ln(t)$$

$$\Delta A_{SWCNT@BNNT} \propto 0.001 \cdot \ln(t)$$

Whereas $\Delta A_{SWCNT}$, $\Delta A_{SWCNT@BNNT}$ represents the changes in absorbance of the SWNT and the SWCNT@BNNT respectively, $t$ is time in second.

According to the fitting results, under an optical power intensity of 13 kW/cm$^2$, the SWCNT@BNNT possesses 48 times longer lifetime than that of the SWCNT for a degradation of 0.5% in absorbance. For 1% degradation, the lifetime is expected to 2,269 times longer.

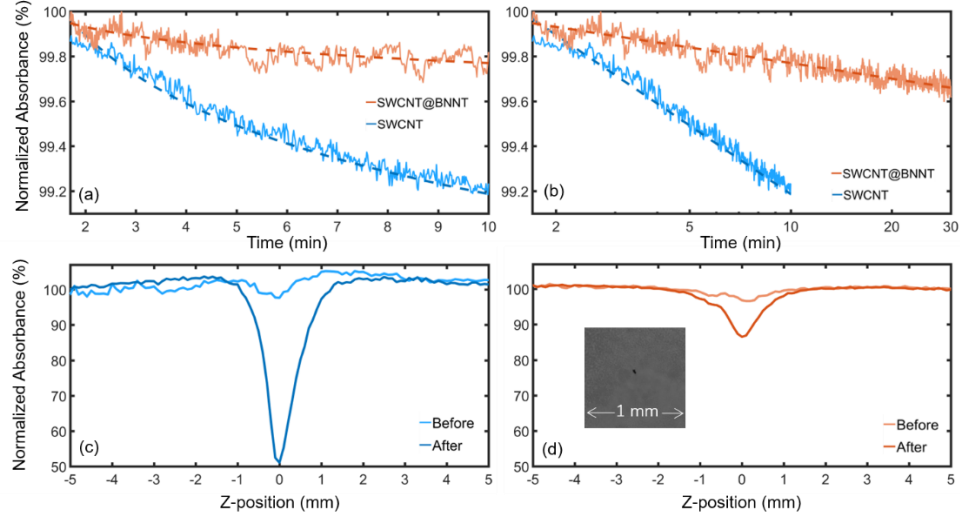

**Fig. 5.** Change of absorbance of pristine SWCNT (blue lines) and SWCNT@BNNT (orange lines) under high optical intensity in (a) linear scale and (b) log scale, and Z-scan result of (c) pristine SWCNT and (d) SWCNT@BNNT before and after heating (inset: picture of damage spot on SWCNT@BNNT under microscope).

After the 10 minutes of illumination, we continue to increase the illumination average power by 5 mW step increment for every 10 minutes from 30 mW up to maximum of 45 mW. The samples are expected to be fully damaged after 40 minutes of illumination with an increasing power, a picture of damaged spot under microscope is shown in inset of Fig. 5 (a). To compare the property of the samples to that before the illumination, a Z-scan measurement is then performed again with a low average power of 5 mW. According to the result shown in Fig. 5 (c, d), both the pristine SWCNT and the SWCNT@BNNT have suffered an irreversible optical damage. As only a small spot on sample is damaged, the overall transmittance measured with larger beam sizes, away from the focal point, is unchanged. However, while the sample is moved in the z-direction closer to the focal point, with smaller beam sizes, more optical power can then pass through the damaged spot, resulting in a significant increase in transmittance. Fig. 5 (c, d) also reveal that after 40 minutes of illumination with an increasing power, the absorbance of the SWCNT and the SWCNT@BNNT is reduced by 50% and 14%, respectively. The higher optical intensity generated more heat accumulated on the sample, resulting in higher degradation in absorbance >10% in tens of minutes, a rate much faster than the estimation made with an average power of 30 mW. These results indicate that the SWCNT@BNNT has a much lower degradation rate in absorbance, and suffers less optical damage at high optical intensities. In order to confirm that the optical damage observed is indeed caused by the heat generated by the focused laser beam, but not the ablation effect of the high peak power of the ultrashort mode-locked pulses, the same experiment has also been carried out using a CW laser source and similar results are obtained.

## 5. Laser Mode-locking using SWCNT@BNNT as SA

These experimental results indicated that BNNT coating on SWCTNT preserves its saturable absorption properties whilst enhancing the thermal tolerance and optical damage threshold. We expect that the SWCNT@BNNT is highly potential to function as a better SA than the pristine SWCNT for laser mode-locking. In order to demonstrate that, a short cavity mode-locking fiber laser using SWCNT@BNNT as an SA has been built for experimental verification. Such a short Fabry-Perot cavity requires much more pump power to generate sufficient gain for laser oscillation, so the SA layer will inevitably be illuminated by an intense pump light. The SA needs to be tolerant to the intense heat generated by both the oscillating signal light and the residue pump light. Pristine CNT is fragile in such a cavity with intense pump light. One way to mitigate that is to use highly-doped Er:Yb phosphor-silicate fibers with a high pump absorption [31, 32]. For SWCNT@BNNT, the high optical damage threshold and heat tolerance allow the use of standard EDF where a high pump power is required and pump leakage is present.

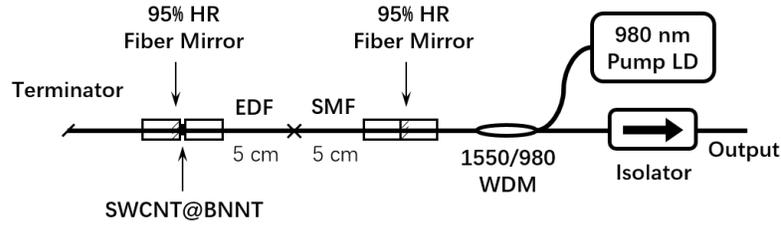

**Fig. 6.** Schematics of the short-cavity mode-locked fiber laser. LD: laser diode, WDM: wavelength division multiplexer, HR: high reflectivity mirror, SMF: single-mode fiber, EDF: Erbium-doped fiber.

The design of the short-cavity mode-locked laser is shown in Fig. 6. The laser cavity consists of a short 10-cm-long fiber cavity sandwiched between two 1550-nm highly reflective (HR) fiber-ferrule mirrors. The HR fiber-ferrule mirror is dichroic, with a 95% reflectivity at 1550 nm wavelength region and transparent at the 980 nm pump wavelength. A wavelength-division multiplexing (WDM) coupler is used to couple the 980nm light from the pump laser diode to the laser cavity. The laser cavity is comprised of a piece of 5-cm-long single-mode fiber (SMF) fusion spliced with a 5-cm-long highly-doped Erbium-doped fiber (EDF). The SMF in the cavity provides an anomalous dispersion ($\beta_2 = -23$ ps$^2$/km) to balance the normal dispersion ($\beta_2 = 30$ ps$^2$/km) of the EDF, resulting in a net cavity dispersion of $3.5 \times 10^{-4}$ ps$^2$. A terminator is employed on the other side of the cavity to prevent undesirable back reflections. On the third port of WDM, an isolator is used at the other port of WDM for laser output.

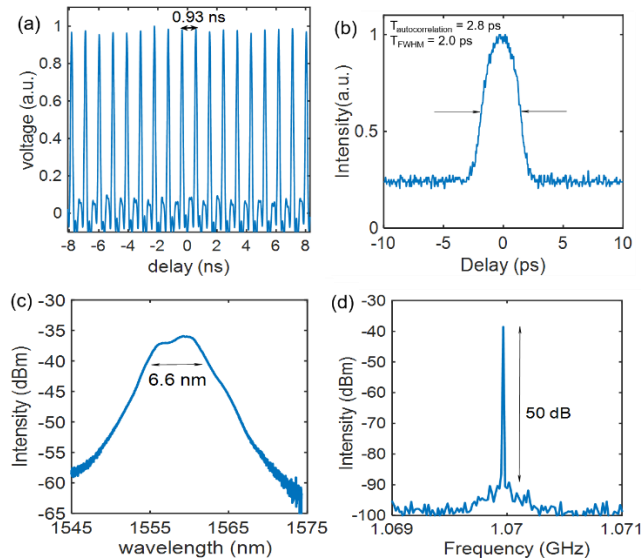

**Fig. 7.** Output characteristics of the fiber laser at a pump power of 470 mW. (a) Oscilloscope trace, (b) Autocorrelation trace, (c) Optical spectrum, and (d) RF spectrum.

Mode-locking can be initiated in the laser cavity using the SWCNT@BNNT as the SA with a pump power of ~180 mW. The average output power is measured to be 50 µW at pump power of 470mW, and the total cavity optical power including the pump- and the signal-light is estimated to be ~ 320 mW. When the pristine SWCNT film is used in the same cavity, the fiber ferrules and the film sandwich get burned out at a pump power level of ~100 mW, well before mode-locking can be initiated.

The oscilloscope trace, autocorrelation trace, output spectrum, and RF spectrum are shown in Fig.7 (a-d) respectively. As the cavity power is too high, multi-pulse was generated in the cavity, so small pulse can be observed on the oscilloscope. The autocorrelation trace in Fig.7 (b) shows an inferred pulse width (FWHM) of 2.0 ps, assuming a Gaussian-pulse shape. Note that, due to the lower level of the laser output, the autocorrelation measurement required a large averaging factor of 256 and the drift in pulse trace location causes an unnatural broadening of the autocorrelation width. We expect the actual pulse width to be narrower than that inferred by the imperfection autocorrelation measurement. According to Fig.7 (c), the full width at half maximum (FWHM) of spectrum is 6.6 nm, and the shape of it indicates that the pulses have been stretched. No side bands and ripple has been observed on the spectrum. The RF spectrum shown in Fig.7 (d) suggests that the repetition rate of cavity is 1.07 GHz and the signal-to-noise ratio is 50dB.

## 6. Conclusion

In conclusion, we investigated the optical properties of a novel nanomaterial named SWCNT@BNNT with novel 1D vdW heterostructure. Its optical saturable absorption properties and optical damage threshold are measured and compared to that of the pristine SWCNT. The results suggest that BN coating on SWCNT can preserve the saturable absorption properties whilst greatly improving the heat tolerance and optical damage threshold. To demonstrate the advantage, a 1 GHz short-cavity mode-locked fiber laser was demonstrated employing SWCNT@BNNT as an SA withstanding an intense optical pump power. This is the first demonstration of using 1D vdW heterostructures to enhance the optical damage threshold of 1D nanomaterials. We believe our work could open up many highly promising possibilities for various nanomaterial-related applications with this new family of nanomaterials with 1D vdW heterostructure. It is expected that this 1D vdW heterostructure technique to be widely applied in profuse optical applications and broaden the horizon of many other fields.

**Acknowledgment**

This work was supported by JSPS Grant-in-Aid for Scientific Research (S) number 18H05238.

**Disclosures**

The authors declare no conflicts of interest.

**References**

1. S. Iijima and T. Ichihashi, "Single-shell carbon nanotubes of 1-nm diameter," nature, **363**(6430), 603-605 (1993).
2. M. F. Yu, O. Lourie, M. J. Dyer, K. Moloni, T. F. Kelly, and R. S. Ruoff, (2000). "Strength and breaking mechanism of multiwalled carbon nanotubes under tensile load," Science, **287**(5453), 637-640 (1993).
3. S. J. Tans, M. H. Devoret, H. Dai, A. Thess, R. E. Smalley, L. J. Geerligs, and Dekker, C. "Individual single-wall carbon nanotubes as quantum wires," Nature, **386**(6624), 474-477 (1997).
4. J. W. Wilder, L. C. Venema, A. G. Rinzler, R. E. Smalley, and C. Dekker, "Electronic structure of atomically resolved carbon nanotubes," Nature, **391**(6662), 59-62 (1998).
5. P. Kim, L. Shi, A. Majumdar, and P. L. McEuen, "Thermal transport measurements of individual multiwalled nanotubes," Physical review letters, **87**(21), 215502 (2001).

6. E. Kymakis and G. A. J. Amaratunga, "Single-wall carbon nanotube/conjugated polymer photovoltaic devices," Applied Physics Letters, **80**(1), 112-114 (2002).
7. M. Freitag, J. Chen, J. Tersoff, J. C. Tsang, Q. Fu, J. Liu, and P. Avouris, "Mobile ambipolar domain in carbon-nanotube infrared emitters," Physical Review Letters, **93**(7), 076803 (2004).
8. M. Freitag, J. C. Tsang, A. Bol, P. Avouris, D. Yuan, and J. Liu, "Scanning photovoltage microscopy of potential modulations in carbon nanotubes," Applied Physics Letters, **91**(3), 031101 (2007).
9. S. Y. Set, H. Yaguchi, Y. Tanaka, and M. Jablonski, "Ultrafast fiber pulsed lasers incorporating carbon nanotubes," IEEE Journal of selected topics in quantum electronics, **10**(1), 137-146 (2004).
10. S. Yamashita, Y. Inoue, K. Hsu, T. Kotake, H. Yaguchi, D. Tanaka, and S. Y. Set, "5-GHz pulsed fiber Fabry-Pe/spl acute/rot laser mode-locked using carbon nanotubes," IEEE Photonics Technology Letters, **17**(4), 750-752 (2005).
11. K. Kieu and M. Mansuripur, "Femtosecond laser pulse generation with a fiber taper embedded in carbon nanotube/polymer composite," Optics letters, **32**(15), 2242-2244 (2007).
12. K. H. Fong, K. Kikuchi, C. S. Goh, S. Y. Set, R. Grange, M. Haiml, and U. Keller, "Solid-state Er: Yb: glass laser mode-locked by using single-wall carbon nanotube thin film," Optics letters, **32**(1), 38-40 (2007).
13. N. Nishizawa, Y. Seno, K. Sumimura, Y. Sakakibara, E. Itoga, H. Kataura, and K. Itoh, "All-polarization-maintaining Er-doped ultrashort-pulse fiber laser using carbon nanotube saturable absorber," Optics express, **16**(13), 9429-9435 (2008).
14. T. Hasan, Z. Sun, F. Wang, F. Bonaccorso, P. H. Tan, A. G. Rozhin, and A. C. Ferrari, "Nanotube–polymer composites for ultrafast photonics," Advanced Materials, **21**(38-39), 3874-3899 (2009).
15. A. Schmidt, S. Rivier, W. B. Cho, J. H. Yim, S. Y. Choi, S. Lee, and U. Griebner, "Sub-100 fs single-walled carbon nanotube saturable absorber mode-locked Yb-laser operation near 1 μm," Optics Express, **17**(22), 20109-20116 (2009).
16. S. Yamashita, "A tutorial on nonlinear photonic applications of carbon nanotube and graphene," Journal of lightwave technology, **30**(4), 427-447 (2011).
17. S. Yamashita, "Nonlinear optics in carbon nanotube, graphene, and related 2D materials," APL Photonics, **4**(3), 034301 (2019).
18. S. Y. Ryu, K. S. Kim, J. Kim, and S. Kim, "Degradation of optical properties of a film-type single-wall carbon nanotubes saturable absorber (SWNT-SA) with an Er-doped all-fiber laser," Optics express, **20**(12), 12966-12974 (2012).
19. A. Martinez and S. Yamashita, "10 GHz fundamental mode fiber laser using a graphene saturable absorber," Applied Physics Letters, **101**(4), 041118 (2012).
20. A. Martinez, K. Fuse, and S. Yamashita, "Enhanced stability of nitrogen-sealed carbon nanotube saturable absorbers under high-intensity irradiation," Optics Express, **21**(4), 4665-4670 (2013).
21. X. Liu, D. Han, Z. Sun, C. Zeng, H. Lu, D. Mao, and F. Wang, "Versatile multi-wavelength ultrafast fiber laser mode-locked by carbon nanotubes," Scientific reports, **3**, 2718 (2013).
22. A. K. Geim and I. V. Grigorieva, "Van der Waals heterostructures," Nature, **499**(7459), 419-425 (2013).
23. K. S. Novoselov, O. A. Mishchenko, O. A. Carvalho, and A. C. Neto, "2D materials and van der Waals heterostructures," Science, **353**(6298) (2016).
24. R. Xiang, T. Inoue, Y. Zheng, A. Kumamoto, Y. Qian, Y. Sato, and K. Hisama, "One-dimensional van der Waals heterostructures," Science, **367**(6477), 537-542 (2020).
25. Y. Chen, J. Zou, S. J. Campbell, and G. Le Caer, "Boron nitride nanotubes: Pronounced resistance to oxidation," Applied physics letters, **84**(13), 2430-2432 (2004).
26. A. G. Nasibulin, A. Kaskela, K. Mustonen, A. S. Anisimov, V. Ruiz, S. Kivisto, and M. Kauppinen, "Multifunctional free-standing single-walled carbon nanotube films," ACS nano, **5**(4), 3214-3221 (2011).
27. Y. Yao, C. Feng, J. Zhang, and Z. Liu, "'Cloning' of single-walled carbon nanotubes via open-end growth mechanism," Nano Letters, **9**(4), 1673-1677 (2009).
28. N. G. Chopra, R. J. Luyken, K. Cherrey, V. H. Crespi, M. L. Cohen, S. G. Louie, and A. Zettl, "Boron nitride nanotubes," Science, **269**(5226), 966-967 (1995).
29. S. Maruyama, Fullerene and Carbon Nanotube Site, http://www.photon.t.u-tokyo.ac.jp/~maruyama/nanotube.html
30. M. Sheik-Bahae, A. A. Said, T. H. Wei, D. J. Hagan, and E. W. Van Stryland, "Sensitive measurement of optical nonlinearities using a single beam," IEEE journal of quantum electronics, **26**(4), 760-769 (1990).
31. S. Yamashita, T. Yoshida, S. Y. Set, P. Polynkin, and N. Peyghambarian, "Passively mode-locked short-cavity 10GHz Er: Yb-codoped phosphate-fiber laser using carbon nanotubes," In Fiber Lasers IV: Technology, Systems, and Applications (Vol. 6453, p. 64531Y). International Society for Optics and Photonics (2007).
32. A. Martinez and S. Yamashita, "Multi-gigahertz repetition rate passively modelocked fiber lasers using carbon nanotubes," Optics express, **19**(7), 6155-6163 (2011).